\documentstyle[12pt,aasms4]{article}
\def \m{\ifmmode M_\odot\else M$_\odot$\fi}

\def\deg{\ifmmode^\circ\else$^\circ$\fi} 

\begin{document}

\title{NICMOS IMAGING OF THE HR 4796A CIRCUMSTELLAR DISK}

\author{GLENN SCHNEIDER\footnote{Steward Observatory, 
University of Arizona, 933 N. Cherry Ave.  Tucson, AZ 85721; 
gschneider, dhines, rthompson, marcia@as.arizona.edu}, 
BRADFORD A. SMITH\footnote{Institute for Astronomy, 
University of Hawaii, Honolulu, HI 96720; 
brad@mahina.ifa.hawaii.edu, meier@hale.ifa.hawaii.edu}, E. 
E. BECKLIN\footnote{Department of Physics and Astronomy, 
University of California, Los Angeles, CA 90095; 
becklin@sofia.astro.ucla.edu, lowrance@astro.ucla.edu}, 
DAVID W. KOERNER\footnote{University of Pennsylvania, 4N14 
DRL, 209 South 33rd Street, Philadelphia, PA 19104; 
davidk@upenn5.hep.upenn.edu}, ROLAND MEIER$^{2}$, DEAN C. 
HINES$^{1}$, PATRICK J. LOWRANCE$^{3}$, RICHARD J. 
TERRILE\footnote{Jet Propulsion Laboratory, MS 183-503, 
Pasadena, CA 91109; Richard.J.Terrile@jpl.nasa.gov}, RODGER 
I. THOMPSON$^{1}$, AND MARCIA RIEKE$^{1}$}

\begin{abstract}

We report the first near infrared (NIR) imaging of a circumstellar annular 
disk around the young ($\sim$8 Myr), Vega-like star, HR 4796A. NICMOS 
coronagraph observations at 1.1 and 1.6 $\mu$m reveal a ring-like 
symmetrical structure peaking in reflected intensity 1.05\arcsec\ $\pm$ 
0.02\arcsec\ ($\sim$70 AU) from the central A0V star. The ring geometry, 
with an inclination of 73.1\deg\ $\pm$ 1.2\deg\ and a major axis PA of 
26.8\deg\ $\pm$ 0.6\deg\, is in good agreement with recent 12.5 and 20.8 
$\mu$m observations of a truncated disk (Koerner, et al. 1998). The ring 
is resolved with a characteristic width of less than 0.26\arcsec\ (17 AU) 
and appears abruptly truncated at both the inner and outer edges. The 
region of the disk-plane inward of $\sim$60 AU appears to be relatively 
free of scattering material. The integrated flux density of the part of 
the disk that is visible (greater than 0.65\arcsec\ from the star) is found 
to be 7.5 $\pm$ 0.5 mJy and 7.4 $\pm$ 1.2 mJy at 1.1 and 1.6 $\mu$m, 
respectively. Correcting for the unseen area of the ring yields total flux 
densities of 12.8 $\pm$ 1.0 mJy and 12.5 $\pm$ 2.0 mJy, respectively 
(Vega magnitudes = 12.92 $\pm$ 0.08 and 12.35 $\pm$ 0.18). 
The NIR luminosity ratio is evaluated from these
results and ground-based photometry of the star. At these wavelengths 
$L_{\rm disk}(\lambda)/L_{*}(\lambda)$ = 1.4 $\pm$ 0.2 
$\times 10^{-3}$ and 2.4 $\pm$ 0.5 $\times 10^{-3}$, 
giving reasonable agreement between the stellar flux scattered 
in the NIR and that which is absorbed in the visible and 
re-radiated in the thermal infrared. The somewhat red reflectance 
of the disk at these wavelengths implies mean particle sizes in excess of 
several microns, larger than typical interstellar grains. The confinement 
of material to a relatively narrow annular zone implies dynamical 
constraints on the disk particles by one or more as yet unseen bodies.

\end{abstract}
 
 \vfill \eject

\section{INTRODUCTION}

HR 4796A (V = 5.78; d = 67 $\pm$ 3.5 pc as determined by {\it Hipparcos}) is a 
young A0V star (Houk 1982), with an estimated age of 8 $\pm$ 3 Myr 
(Stauffer, et al. 1995, Jayawardhana et al. 1998). It has been the 
object of much scrutiny since Jura (1991) inferred the presence of an 
unusually large amount of circumstellar dust, based on an analysis of IRAS 
data. Jura's initial estimate of the dust opacity, $\tau_{\rm dust} = 
L_{\rm dust}/L_{*} = 5\times 10^{-3}$, exceeds by a factor of two that of 
$\beta$ Pictoris, then the only main sequence star for which a 
circumstellar disk had been optically imaged (Smith \& Terrile 1984). Jura 
et al. (1995) noted that their earlier estimated dust temperature (110K) 
indicated a lack of disk material within 40 AU of HR 4796A. They required a 
minimum grain size of $\sim$ 3 $\mu$m for the particles to remain 
gravitationally bound and suggested that this material is located between 
40 AU and 200 AU from the star. They further suggested that grains of this 
size must have coalesced from smaller particles, with a much higher growth 
rate into larger particles taking place within 40 AU of the star. Their 
2$\mu$m speckle data and optical spectra ruled out any close stellar 
``sweeper'' companions with M$_{*} > 0.125 $M$_\odot$ as close as 11 AU 
that could be responsible for such a central clearing (Artymowicz \& Lubow 
1994), but left open the possibility of lower mass companions.

Recently a circumstellar disk around HR 4796A was imaged
at 12.5 $\mu$m and 20.8 $\mu$m by Koerner et al. (1998) and
observed independently at lower resolution by Jayawardhana
et al. (1998) at 18.2 $\mu$m. An inner depleted region 
was apparent in the high-resolution 20.8$\mu$m image, and this was 
reproduced by a model of the emission from a disk with inclination, i = 
72\deg\ (+6\deg, -9\deg), position angle 28\deg\ $\pm$ 6\deg, and inner 
radius, R$_{in}$ $\sim$ 55 AU. Additional constraints from long-wavelength 
flux densities yielded an outer radius, R$_{out}$ $\sim$80 AU. In 
contrast, the excess 12.5$\mu$m emission appeared centered on the star and, 
together with small residuals from the fit at 20.8$\mu$m, indicated the 
presence of a tenuous warm dust component confined to within a few AU of 
the star. These results were interpreted by Koerner et al. (1998) to 
indicate the presence of an inner compact ``zodiacal dust'' component 
surrounded by a more prominent dust ring at a radial distance corresponding 
to the Kuiper Belt.

Ground-based attempts to image the HR 4796A disk in reflected light, (e.g., 
Mouillet, et al. 1997), have thus far failed. However, NICMOS 1.1$\mu$m 
and 1.6$\mu$m images of a ring-like circumstellar annular disk about 
HR 4796A now provide for the first time spatially resolved photometric 
properties of the disk, and more precisely define its morphology. A 
first-look analysis of the data is reported here. Detailed modeling of the 
geometric and photometric properties of the disk will be discussed in a 
later paper.

\section{NICMOS CORONAGRAPHIC OBSERVATIONS}

NICMOS Camera 2 coronagraphic observations of HR 4796A were obtained at two 
epochs, 1998 March 15 and 1998 August 16. The initial observations, in 
which both the NE and SW ansae of the annulus were detected, were obtained 
in H-band (F160W filter; $\lambda_{\rm central} = 1.594\mu$m, 
$\Delta\lambda = 0.403\mu$m). The second-epoch follow-up observations 
employed both the F160W and F110W ($\lambda_{\rm central} = 1.0998 \mu$m, 
$\Delta\lambda = 0.592 \mu$m) filters. Although the geometrical radius of 
the coronagraph hole is 0.3\arcsec, instrumental scatter, diffraction and 
Point Spread Function (PSF) mis-registration extended the radius of 
non-usable data to $\sim$ 0.65\arcsec. To further facilitate 
discrimination of PSF artifacts (which are more prevalent at 1.6$\mu$m) 
from intrinsic morphological features of the circumstellar disk, the 
rotational orientations of the field at each epoch differed by 168.8\deg.

By obtaining coronagraphic images at different field orientations (by 
rotating the spacecraft about the target axis) it is possible to 
differentiate unocculted circumstellar objects and features from artifacts 
caused by instrumental scattering and the complex {\it HST}+NICMOS pupil 
function. Optical/instrumental artifacts co-rotate with the detector, 
while objects/features in the target image do not. 
The detection of faint point-like objects, or low-surface-brightness 
extended features of angular azimuthal extent smaller than their position 
displacements due to field rotation, is made possible by subtracting
the second roll orientation image from the first thereby nulling the
unocculted wings of the occulted bright target PSF.
This works extremely well to the level of the residual 
photon noise if the target is re-positioned in the coronagraph hole with 
high precision following a spacecraft roll (Schneider et al. 1998).
Individual sets of images were obtained within 25 minute
intervals to minimize PSF variations that arise from ``breathing'' of the 
{\it HST} optical telescope assembly (Bely, 1993). To eliminate 
image artifacts known as ``the bars'' 
(STScI 1997), Cameras 1 and 3 were run simultaneously (but blanked off in 
ACCUM mode) during all of the Camera 2 observations.

The 1998 March 15 (first-epoch) observations were carried out using an 
observing strategy described by Lowrance, et al (1998a), and adopted for 
the search/detection phases of the NICMOS Environments of Nearby Stars 
(EONS) observing programs (Schneider 1998). Imaging sequences were 
executed with a total integration time of 672s at each of two image 
orientation angles differing by 29.9\deg. However, data from the second 
orientation were significantly degraded due to a guide star acquisition 
failure and could not be used.  An available reference PSF from unrelated 
observations of CoD -33\deg\ 7795 
(Lowrance et al. 1998b) was subtracted from the data. 
The difference image indicated the presence of excess flux, spatially 
coincident with the locations of the brightest parts of the 20.8$\mu$m disk 
found by Koerner, et al. (1998). Based on this initial 1.6$\mu$m 
detection, follow-up observations were planned for 16 August 1998 after 
HR 4796A came out of {\it HST} solar avoidance. Later, a ring-like annular 
disk was clearly seen in this first-epoch data after subtracting a better 
matched PSF from another star (51 Oph, see Figure 1a.).

Observations at 1.1$\mu$m were added at the second-epoch to obtain color 
information on the disk. Instrumental scatter at small radial distances 
from the edge of the coronagraph hole is reduced significantly at shorter 
wavelengths. These F110W coronagraphic images were obtained at two 
spacecraft roll orientations, used here as a ``roll dither" to replace bad 
pixels rather than for difference-imaging, source detection. Because of 
sun-angle and available guide-star constraints, a differential roll of only 
9\deg\ was possible. At 1\arcsec\ (where the 1.6$\mu$m excess flux peaked 
in the initial detection image), a 9\deg\ roll results in a field rotation 
of 0.158\arcsec, slightly more than two pixels. This exceeds the FWHM of 
the PSF at 1.1$\mu$m by $\sim$32\%. Thus, beyond the 0.65\arcsec\ inner 
usable radius, roll-paired images may be effectively re-sampled around bad 
pixels. The total integration time of the F110W data was 1952s. Four 
F160W images, with a total integration time of 1024s, were obtained at a 
single field orientation that differed from that of the 15 March 
observations.

At the second roll position the spacecraft was slewed to a nearby bright 
star, HR 4748 ({\it H} = 5.44) of similar spectral type (B8V) to produce a 
contemporaneous F110W coronagraphic reference PSF in the same orbit. This 
star exhibited no evidence of multiplicity in previous ground and 
space-based observations. A total integration time of 576s was used to 
achieve S/N comparable to that obtained for HR 4796A.

\section{DATA CALIBRATION AND REDUCTION}

The raw MULTIACCUM image data were calibrated with an analog to STScI's 
CALNICA pipeline software, with three differences in processing: (1) Before 
the standard linearity corrections, flux estimates were made for pixels 
that were driven into non-linearity in either the first and second reads of 
the detector following the initial reset. (2) Cosmic ray hits were 
detected (and subsequently compensated for) with a sigma-clipped cosmic-ray 
rejection procedure. (3) Corrections were made for residual DC pedestals 
and quadrant-dependent DC offsets in the detector. These steps were taken 
before flat-fielding the derived count-rate image. Calibration dark 
reference files made from on-orbit darks by the NICMOS IDT were used. 
Standard Camera 2 bad-pixel maps were updated for changes in 
under-responsive pixels from contemporaneous, background-subtracted, F160W 
flat-fields obtained as part of the target acquisition (TA) process. In 
the case of the F160W imaging, the reference flat-fields themselves were 
augmented near the edge of the coronagraph hole with the TA flats to better 
compensate for short-term drifts of the hole position. While TA flats are 
inherently of lower S/N ($\sim$100) than standard calibration flats (S/N 
$\sim$1200), this is more than sufficient to remove those artificial 
edge-gradients that are otherwise introduced by the application of 
uncorrected flat-fields. Because translational dithering is generally not 
feasible in coronagraphic observations, bad pixels in the flux-calibrated 
frames were replaced by 2D Gaussian-weighted interpolation. A weighting 
radius of 3 pixels was used for the F110W data and 5 pixels for the F160W 
data. After bad-pixel replacement the fluxes derived from the calibrated 
MULTIACCUMs were averaged with equal weighting. This process was applied 
to the HR 4796A (F110W and F160W), HR 4748 (F110W) and 51 Oph (F160W) 
MULTIACCUM images.

The HR 4748 and 51 Oph calibrated reference PSFs were subtracted from the 
HR 4976A F110W and first-epoch F160W calibrated images using the {\it idp3} 
program developed by the NICMOS IDT (Lytle et al., 1998). The reference PSFs were 
rescaled in flux to match the stellar component of the HR 4796A images. 
Re-registration was accomplished through bi-cubic convolution interpolation 
(Park \& Schowengerdt 1983). The residual energy along the three-component 
diffraction spikes was measured (at radii where the circumstellar source 
flux is not contributing significantly) and effectively removed (Fig. 1b) 
in the F110W PSF-subtracted image.

\section{GEOMETRY, MORPHOLOGY AND PHOTOMETRY}

a) GEOMETRY: Implicit in this analysis is the assumption that the observed 
circumstellar ring structure is circular. The orientation, inclination, 
and mean radius of the circumstellar ring were found, independently, from 
the first and second epoch F160W (single orientation), and second epoch 
F110W (combined two-orientation) images by least-squares ellipse fitting of 
the ``mid-ring'' and the inner and outer half-peak intensity contours. The 
solutions from each data set were constrained such that the resulting fits 
to the isophotes were required to be concentric. In all three cases the 
parametric solutions overlapped within their errors. The solution from the 
F110W observations, which have the highest statistical significance, has 
been adopted. These observations are of higher S/N and are less biased by 
the influence of residual PSF artifacts than those of the F160W 
observations. The solution yields a position angle of 26.8\deg\ $\pm$ 
0.6\deg, an inclination of 73.1\deg\ $\pm$ 1.2\deg\ and a semi-major axis of 
1.05\arcsec\ $\pm$ 0.02\arcsec\ for the projected ring ellipse. Assuming a 
distance to HR 4796 of $\sim$67 pc, this corresponds to a physical radius of 
70.4 $\pm$ 1.4 AU, in excellent agreement with Koerner, et al. (1998) 
(PA = $28\deg\pm6\deg$, i~=~72\deg\ (+6\deg, -9\deg), r = 67.5 AU $\pm$ 12.5 
AU). In the F110W data no statistically significant deviations from 
ellipticity were found. The ellipse fits are superimposed on 1.6 and 1.1 
$\mu$m isophotal maps of the HR 4796A disk in Fig. 1c and 1d. It is 
important to emphasize here that these solutions are based solely on 
observed isophotes and do not, as yet, consider the scattering properties 
of the disk material.

b) MORPHOLOGY: The ring appears to be abruptly bounded both interior and 
exterior to the mid-ring (r $\sim$ 70AU) in both the 1.1 and 1.6$\mu$m 
images. At 1.1$\mu$m the FWHM of the mean radial component of the peak 
intensities of the ansae is less than 0.29\arcsec\ ($\sim$20 AU). The 
measured profile is broadened by the coronagraphic PSF, which itself has a 
FWHM of 0.12\arcsec\ (8 AU at the distance of HR 4796). Quadratic 
subtraction of the PSF yields an actual photometric full-width of the disk 
annulus equal to or less than 0.26\arcsec\ (17 AU). No significant level 
of scattered radiation is seen inward from $\sim$60 AU to $\sim$45 AU where 
the residual photon noise from the bright central star dominates the 
measurements.

c) PHOTOMETRY: Figure 1 (c\&d) shows the measured flux isophotes of the 
HR 4796A disk at both 1.1µm and 1.6$\mu$m. Photometric calibration 
constants of 2.195$\times 10^{-6}$ and 2.207$\times 10^{-6}$ Jy ADU$^{-1}$ 
sec$^{-1}$ (Rieke 1998) were used to convert from instrumental count rates 
to physical units for the F110W and F160W filters, respectively.

The total reflected energy outside a radius of 0.65\arcsec\ was estimated 
by summing the flux within a rectangular region of $\sim3.1\arcsec \times 
0.9\arcsec$ (corresponding to the axial ratio of the disk, but oversized by 
50\%) oriented along the disk major axis. In doing so the residual 
background from four nearby regions adjacent to this area was subtracted 
and the photometric error, as determined from the background variations, 
was estimated. In the F160W image, spurious PSF artifacts were removed by 
treating them as regions of ``missing data'' and then interpolating 
bi-cubically over the small areas. The total flux measured was 7.6 $\pm$ 
0.5 mJy at 1.1 $\mu$m and 7.4 $\pm$ 1.2 mJy at 1.6$\mu$m. The area of 
ring contained within the inner 0.65\arcsec\ radius region, and thus not 
measurable, is $\sim$ 40.8\% of the total area. Applying an areal 
correction (and assuming uniform surface brightness), the total flux is 
found to be 12.8 $\pm$ 1.0 mJy at 1.1$\mu$m and 12.5 $\pm$ 2.0 mJy at 
1.6 $\mu$m. Zero-points for the photometric conversions, also from Rieke 
(1998), are 1909 and 1087 Jy for an H=0.0 source for the F110W and F160W 
filters, respectively. These give integrated Vega magnitudes for the ring 
of 12.92 $\pm$ 0.08 (F110W) and 12.35 (-0.19, +0.16) (F160W). Using 
ground-based multi-color photometry of HR 4796A (Jura, et al. 1993), 
({\it I} = 5.81, {\it J} = 5.80, {\it H}= 5.80, all $\pm$ 0.10), the 
NIR optical 
depth is $\tau_{\rm NIR} \sim L_{\rm disk}(\lambda)/L_{*}(\lambda) = 1.4 
\pm 0.2 \times 10^{-3}$ and $2.4 \pm 0.5 \times 10^{-3}$, respectively. 
Thus, there is first-order agreement between
the stellar flux scattered by the circumstellar dust in the
NIR and that which is absorbed in the visible and re-radiated
by the dust in the thermal infrared
($L_{\rm disk}/L_{*} = 5 \times 10^{-3}$). 
Within the photometric errors, the ring appears to be somewhat 
red in reflection. This suggests that the mean 
disk-particle size must be larger than several microns, indicating 
debris origin rather than trapped interstellar dust.

\section{SUMMARY AND CONCLUSIONS}
NICMOS 1.1$\mu$m (F110W) and 1.6$\mu$m (F160W) coronagraphic observations 
have for the first time imaged the HR 4796A circumstellar disk in reflected 
light, revealing a narrow, ring-like structure. These images provide 
spatially resolved geometric and photometric properties of the disk using 
the high angular resolution offered by {\it HST}. The ring, with a mean radius 
of approximately 70 AU, appears sharply bounded and narrow ($<$ 17 AU). 
The NICMOS observations effectively probe the circumstellar environment 
inward to a radius of $\sim$ 45 AU from the central star and confirm the 
depletion of disk material inferred by earlier observers (e.g., Jura et al. 
1993) and, more recently, by direct observations at 20.8 $\mu$m (Koerner et 
al. 1998). The observed flux ratio, $L_{\rm disk}(\lambda)/L_{*}(\lambda) 
= 1.4 \pm 0.2 \times 10^{-3}$ at 1.1 $\mu$m and $2.4 \pm 0.5 \times 
10^{-3}$ at 1.6 $\mu$m, indicating somewhat red reflection. 
This implys a mean particle size that is large compared to these 
wavelengths. Assuming $\tau_{\rm dust} \sim L_{\rm 
disk}(\lambda)/L_{*}(\lambda)$, there is reasonable agreement between the 
scattered stellar flux in the NIR and that which is absorbed in the visible
and re-radiated in the thermal infrared. The stable containment 
of ring particles and the abrupt inner
and outer truncation of the ring around this relatively young star
imply dynamical constraints (Goldreich \& Tremaine, 1979) imposed
by one or more related, but as yet unseen, bodies.  How such bodies,
presumably planets, could have formed so quickly at such large
distances from the star presents an interesting challenge to those
who model planet formation.

\acknowledgments {We thank our program coordinator, 
Douglas Van Orsow, and contact scientist, Alfred Schutz, at STScI for the 
assistance in implementing and scheduling our {\it HST}/EONS observations. 
This work is supported by NASA grant NAG 5-3042. This paper is based on 
observations with the NASA/ESA {\it Hubble Space Telescope}, obtained at 
the Space Telescope Science Institute, which is operated by the Association 
of Universities for Research in Astronomy, Inc. under NASA contract 
NAS5-26555.}
 

\newpage
\centerline{\bf FIGURE CAPTIONS}
\par\noindent

Fig 1 -- Images and 10 $\mu$Jy isophotal maps of the HR 4796A 
circumstellar ring. Images a\&c: 1.6 $\mu$m (F160W), 15 March 1998. 
Images b\&d: 1.1 $\mu$m (F110W), 16 August 1998. The unusable area 
circumscribing the coronagraphic hole is indicated (gray circles). A 
few artifacts in the 1.6$\mu$m image remain at larger radii arising 
from an imperfectly matched stellar PSF, and are identified in image 
c (red arrows). These were removed in image a
based on comparison of this PSF subtraction with the 
second-epoch F160W images and a priori knowledge of the NICMOS 
coronagraphic pupil function. The coronagraph and detector were 
rotated 168.8\deg\ between observational epochs, aiding in the 
identification (and rejection) of optical artifacts.
Color bars give flux densities in $\mu$Jy per Camera 2 pixel 
(0.0762\arcsec\ x 0.0755\arcsec\ both $\pm$ 0.0002\arcsec).

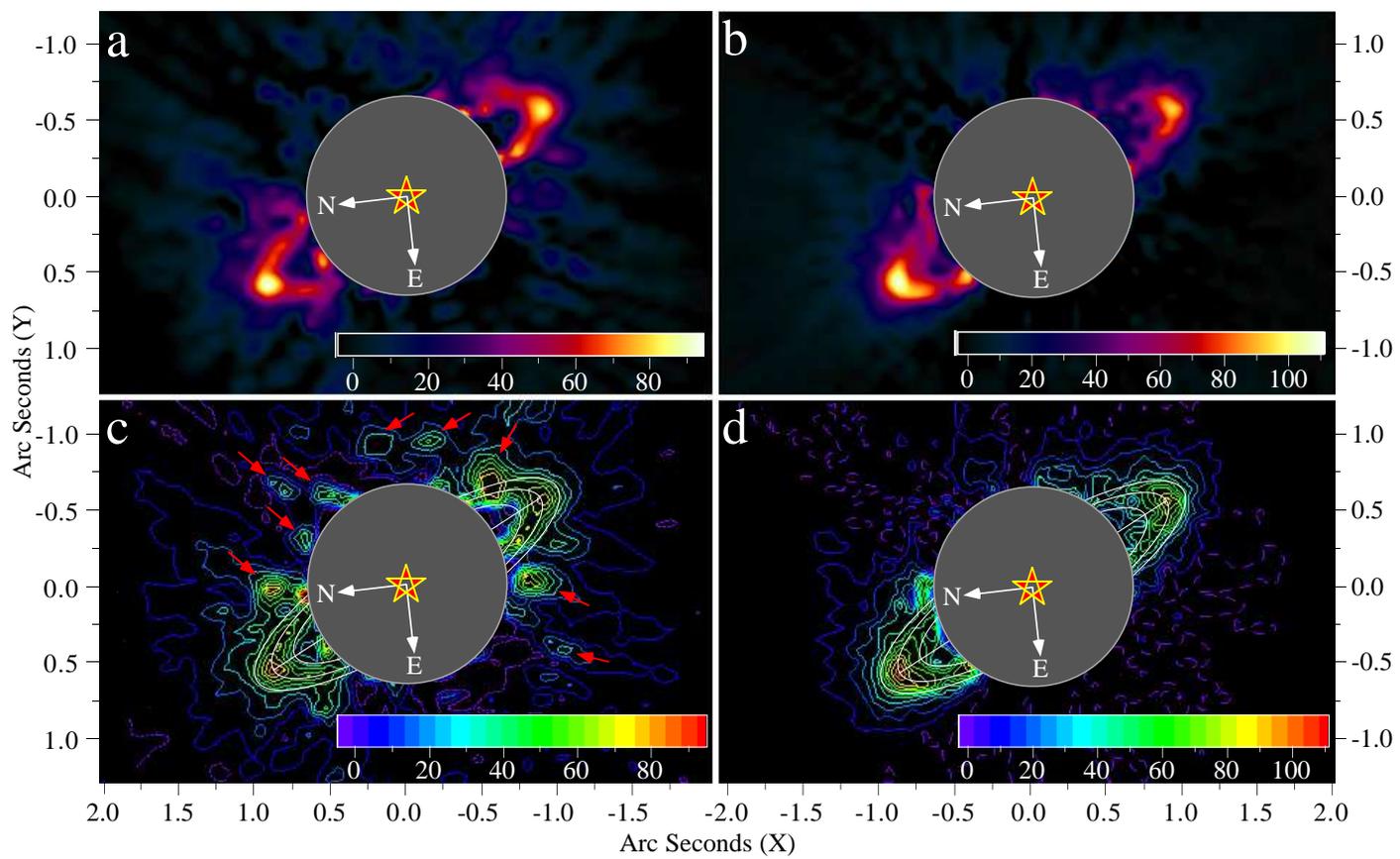
\end{document}